\documentstyle[11pt,aaspp4]{article}
\begin{document}
\title{HEAVY ELEMENTS IN QSOS:\\
STAR FORMATION AND GALAXY EVOLUTION\\ AT HIGH REDSHIFTS}

\author{F. Hamann} 
\affil{University of California, San Diego, USA} 

\begin{abstract}\small
Intrinsic emission and absorption lines of QSOs provide 
several independent probes of the metal abundances in QSO 
environments. They indicate that the metallicities are typically 
solar or higher out to redshifts $z>4$. These 
results support models of galaxy evolution where galactic 
nuclei, or dense condensations that later become galactic 
nuclei, form stars and evolve quickly at redshifts higher 
than the QSOs themselves. 
\end{abstract}

\section{Introduction}

Measuring the heavy-element abundances near QSOs can provide 
unique constraints on high-redshift star formation and galaxy 
evolution. In particular, QSO abundances reflect the 
evolution characteristics of galactic nuclei or dense 
proto-galactic condensations at high redshifts -- 
perhaps involving the first generations of 
stars formed after the Big Bang. The QSO results will therefore 
complement other studies of high-redshift 
galaxies that probe more extended structures and/or rely on very 
different data and analyses techniques. Combining the QSO abundance 
work with the other studies, involving, for example, 
the ``Lyman-break'' galaxies and damped-Ly$\alpha$ absorbers, 
should yield a more complete picture of star formation and galaxy 
evolution at early epochs. 
\vskip 0.2cm

Three general, independent probes of QSO abundances are 
readily observable at all redshifts: the broad emission lines (BELs), 
the broad absorption lines (BALs) and the intrinsic narrow 
absorption lines (NALs). Each of these probes has its own 
theoretical and observational uncertainties, so it is 
essential to consider as many of them as possible. I am now involved in 
several projects to examine a wide range of abundance 
diagnostics in QSOs at different redshifts and luminosities. 
My principle collaborators are Drs. T. Barlow, F. Chaffee, 
G. Ferland, C. Foltz, V. Junkkarinen, K. Korista and J. Shields. 

\section{Results}

\noindent\underline{\bf BELs.} \ \ 
The broad emission lines have a major advantage in that they 
can be measured and compared in large samples of QSOs using 
using moderate resolution spectra. Line ratios 
involving nitrogen are particularly valuable as tracers of the 
chemical enrichment because N is selectively enhanced by 
``secondary'' processing in stellar populations 
(increasing roughly as $Z^2$ for at least $Z\ga 0.2 Z_{\odot}$; 
see \cite{hf93},\cite{v93}). 
Hamann \& Ferland (\cite{hf92},\cite{hf93},\cite{f96}) 
used extensive photoionization calculations to show that the 
broad emission-line ratio NV $\lambda$1240/HeII $\lambda$1640 provides 
lower limits on the N/He abundances. This abundance sensitivity 
occurs because the NV and HeII lines form together within the 
He$^{++}$ zone; NV can be relatively weak for 
some nebular parameters (e.g. gas density, ionizing flux and 
continuum shape, etc.), but it is not possible to produce large  
NV/HeII ratios without increasing N/He. 
The same calculations show that NV/CIV $\lambda$1549 
can be an indicator of N/C. 
\vskip 0.2cm

Applying this analysis to observed NV/HeII and NV/CIV ratios in 
QSOs (see also \cite{h97c}), together with simple 
considerations from galactic chemical evoltuion, indicates 
that 1) QSOs out to $z>4$ typically have solar or higher 
metallicities, 2) the stellar initial mass function favors 
massive stars (slightly) more than in the solar neighborhood, and 
3) the enrichment timescales are
$\la$1~Gyr for at least the $z>4$ sources (if $\Omega_o$ $\approx$ 1).
Similar evolution characteristics have been inferred from 
the old stellar populations in present-day elliptical galaxies and 
spiral bulges; vigorous star formation in those environments should 
produce gas-phase abundances above solar at early cosmological 
epochs. (The present-day stellar populations represent 
an integral over all previous gas-phase abundances and therefore 
have lower {\it average} metallicities.) 
There is also a trend in the NV line ratios suggesting that more 
luminous QSOs have higher metal abundances (\cite{hf93}, 
\cite{o94}). If QSO luminosities 
are tied to the mass of their host galaxies (eg. \cite{m95}), 
this tentative luminosity-metallicity trend could derive from a 
mass-metallicity relation among QSO hosts that is analogous 
(or identical) to the well-known relationship in nearby galaxies. 
\vskip 0.2cm

\noindent\underline{\bf NALs.} \ \ 
Abundance estimates from QSO absorption lines are, in principle, 
more straightforward than the emission lines because the 
line strengths are not sensitive to the gas 
densities or temperatures. Moreover, absorption 
lines yield direct measures of the column densities in different 
ions. One has only to apply appropriate ionization corrections 
to convert the column densities into relative abundances. 
For example, the abundance ratio for any two elements $a$ and 
$b$ can be written as, 
\begin{equation}
\left[{a\over b}\right] \ = \ \ 
\log\left({{N(a_i)}\over{N(b_j)}}\right) \ +\
\log\left({{f(b_j)}\over{f(a_i)}}\right) \ +\
\log\left({b\over a}\right)_{\odot} 
\end{equation}
where $(b/a)_{\odot}$ is the solar abundance ratio, and $N$ and $f$ 
are  respectively the column densities and ionization fractions 
of elements $a$ and $b$ in ion states $i$ and $j$. 
Ideally, one has abundance-independent constraints on 
the ionization fractions from the column densities of different 
ions of the same element. Otherwise, we can also constrain 
the ionization by comparing column densities in different 
elements, with some assumption about their relative abundance. 
If there is a range of ionization states or a complete 
lack measured constraints (eg. if only HI and CIV lines are 
measured), it is still possible to use 
minimum ionization correction factors to derive 
minimum metal-to-hydrogen abundance ratios. Hamann (\cite{h97a}) 
plotted theoretical correction factors for a wide variety of 
circumstances. 
\vskip 0.2cm

A central issue in using NALs for abundance work is understanding 
the location of the absorbing gas. Some of the so-called 
``associated'' (or $z_a\approx z_e$) absorbers 
might reside very near the QSOs, perhaps in outflows similar 
to BALs, but others could form in the extended halo of the host 
galaxy or in cosmologically intervening gas. 
Every system must be examined 
individually. Several empirical tests have been developed to 
help identify NAL systems that are truly intrinsic to QSO 
environments, for example 1) 
time-variable line strengths, 2) multiplet ratios that imply 
partial line-of-sight coverage of the background light 
source(s), and 3) well-resolved line profiles that are 
smooth and broad compared to thermal line widths 
(see \cite{h97b}, \cite{b97} and references therein). 
\vskip 0.2cm

I am in the midst of a program to identify intrinsic NALs 
and measure there abundances. In all three bonafide intrinsic 
systems studied so far, the metallicity is solar or higher 
(see \cite{h97b}, \cite{h97c}). This result agrees with 
the few other known cases of intrinsic NALs (\cite{p94}, \cite{w93}) 
and with the preponderance of $Z\ga $~Z$_{\odot}$ results for 
general $z_a\approx z_e$ systems of uncertain origin 
(\cite{h97a},\cite{tr97},\cite{s98},\cite{s94}). 
\vskip 0.2cm

\noindent\underline{\bf BALs.} \ \ 
The most surprising abundance results have come from studies 
of the BALs, where the strengths of the metal-lines 
compared to HI Ly$\alpha$ seem to require metallicities (for 
example Si/H) from 20 to $>$100 times solar 
(\cite{t96},\cite{h97a}). The secure 
detections of broad PV~$\lambda\lambda$1118,1128 absorption 
in two BALQSOs (\cite{j97},\cite{h98}), 
and tentative detections in two others,  
suggest further that  phosphorus is highly 
overabundant, with P/C~$>$~60 and P/H~$>$1000 times solar  
(see also \cite{h97a}). These abundances, particularly the high 
P/C, are not only in conflict with the other diagnostics but 
they are also incompatible with any enrichment scheme 
dominated by Types I or II supernovae or CNO-processed material 
from stellar envelopes. 
\vskip 0.2cm

Another possibility is that the BAL abundance estimates 
are simply incorrect. 
Hamann (\cite{h98}) argued that the PV BAL has a significant 
strength {\it not} because phosphorus is overabundant, but 
because the strong transitions like CIV, NV and OVI are 
much more optically thick than they appear. 
Explicit calculations of the line optical depths assuming solar 
relative abundances show that PV is the first weak line to 
appear as the stronger transitions become 
more saturated. The strength of the PV BAL in PG~1254+047 
implies optical depths of, for example, $\ga$6 in 
Ly$\alpha$, $\ga$25 
in CIV and $\ga$80 in OVI for solar relative abundances. 
These results indicate that the column densities derived from 
the measured troughs are gross underestimates and, 
consequently, the true abundances are unknown. 
\vskip 0.2cm

BALs like CIV, OVI and Ly$\alpha$ might be optically thick while 
not reaching zero intensity 
if the absorber covers just part of the continuum source(s).
Furthermore, different optically thick lines 
can have different strengths and profiles if their coverage fractions 
differ. Note that coverage fraction differences between BALs 
(that mimic simple 
optical depth or ionization effects in observed spectra) 
can occur naturally if the absorbing regions have a range of 
ionization states or column densities. There is already direct 
evidence for partial coverage, and sometimes 
different coverage fractions in different lines, from the resolved 
multiplet ratios in the narrow components of some BALs 
(\cite{b94}, \cite{w95}) and 
intrinsic NALs (\cite{b97}, \cite{h97a}, \cite{h97b} 
and refs. therein). Partial coverage has also been inferred from 
spectropolarimetry of BALQSOs. We cannot measure the coverage 
fractions from multiplet ratios in 
most BALs, but I claim that the significant strength 
of PV absorption signifies partial  
coverage and large line optical depths. Strong support for 
this interpretation comes from the only known NAL system 
with PV absorption, where the resolved doublets clearly 
indicate large optical depths and partial coverage in lines 
such as CIV, NV and SiIV (\cite{b98}). 
\vskip 0.2cm

Thus, the true BAL abundances are unknown. 
I am presently involved in a program to obtain spectra 
of BALQSOs across a wide range of rest UV wavelengths. 
Our goals are to 1) determine the general strength and frequency 
of PV absorption in BALQSOs, and 2) see if 
some BAL systems or some portions of BAL profiles (the 
high-velocity wings?) might still be useful for abundance 
studies.

\section{Summary and Discussion}

In spite of the null results from BALs, a consensus is emerging 
from the BELs and intrinsic NALs for typically solar 
or higher metallicities in high-redshift QSOs. 
These results support models of galaxy evolution  
wherein vigorous star formation in galactic nuclei, or 
dense proto-galactic condensations, produces super-solar 
gas-phase metallicities at redshifts $>$4. 
High metal abundances are a signature of deep gravitational 
potentials and thus {\it massive} galaxies or proto-galaxies 
because only they can retain their gas long enough 
against the building thermal pressures from supernova explosions. 
The enriched gas might ultimately be ejected from the 
galaxy, consumed by the black hole, or diluted by subsequent infall, 
but the evidence for early-epoch high-$Z$ gas remains in the stars today.  
In particular, the mean stellar metallicities in the cores of nearby 
massive galaxies are typically $\sim$1 to 3~Z$_{\odot}$. The individual 
stars are distributed about these means with metallicities reflecting the 
gas-phase abundance at the time of their formation. Only the most 
recently formed stars at any epoch have metallicities as high as 
that in the gas. Simple chemical evolution models 
indicate that the gas-phase abundances in galactic nuclei 
should be $\sim$2 to 3 times larger than the stellar means, e.g.   
$\sim$2 to 9~Z$_{\odot}$ near the end of the star-forming epoch 
(see \cite{hf93}, \cite{h97a} and references therein). 
Therefore, metallicities in the range $2\la Z\la 9$~Z$_{\odot}$
can be {\it expected} in QSOs as long as considerable 
star formation and enrichment 
occurs before the QSOs ``turn on'' or become observable.

\acknowledgements{This work was supported by NASA grants NAG 5-1630 
and NAG 5-3234}


%
\vfill
\end{document}